\documentclass[jgr]{AGUTeX}

\pdfoutput=1

\pdfoutput=1
\usepackage[dvips,pdftex]{graphicx}
\usepackage[]{caption}
\usepackage{morefloats}
\usepackage{color}
\usepackage[rgb,dvipsnames]{xcolor}
\definecolor{grey}{gray}{0.7}
\usepackage{amsmath}
\usepackage{amssymb}
\usepackage{latexsym}
\usepackage{enumitem}
\usepackage{setspace}
\setkeys{Gin}{draft=false}

\newlength{\myeqlength}
\makeatletter
\g@addto@macro \normalsize {%
  \setlength\abovedisplayskip{\myeqlength}%
  \setlength\belowdisplayskip{\myeqlength}%
}
\makeatother
\makeatletter
\g@addto@macro \normalsize {%
  \setlength\abovedisplayshortskip{\myeqlength}%
  \setlength\belowdisplayshortskip{\myeqlength}%
}
\makeatother
\setlist{leftmargin=3.5mm}

\titlerunninghead{x-line orientation}
\authorrunninghead{Yi-Hsin Liu et al.}

\authoraddr{Yi-Hsin Liu,
NASA Goddard Space Flight Center,
8800 Greenbelt Rd,
Bldg. 21, Room 053c, Code 670,
Greenbelt, MD 20771, USA.
(yhliu10@gmail.com)}

\begin{document}
\bibliographystyle{agu08}

\title{Orientation of x-lines in asymmetric magnetic reconnection -- mass ratio dependency}

 \authors{Yi-Hsin Liu,\altaffilmark{1}
 M. Hesse,\altaffilmark{1} and M. Kuznetsova\altaffilmark{1}}

\altaffiltext{1}{NASA-Goddard Space Flight Center, Greenbelt, Maryland, USA.}

\begin{abstract}
Using fully kinetic simulations, we study the x-line orientation of magnetic reconnection in an asymmetric configuration. A spatially localized perturbation is employed to induce a single x-line, that has sufficient freedom to choose its orientation in three-dimensional systems. The effect of ion to electron mass ratio is investigated, and the x-line appears to bisect the magnetic shear angle across the current sheet in the large mass ratio limit. The orientation can generally be deduced by scanning through corresponding 2D simulations to find the reconnection plane that maximizes the peak reconnection electric field. The deviation from the bisection angle in the lower mass ratio limit can be explained by the physics of tearing instability.   
\end{abstract}

\begin{article}
\section{Introduction}  \label{sec:introduction}
Magnetic reconnection is arguably one of the most important energy conversion and plasma transport processes in solar and space plasmas. Among other effects, it determines the energy entry from the solar wind into Earth's magnetosphere, and it enables energy transport and dissipation therein \citep{dungey61a}. At Earth's magnetopause, reconnection proceeds asymmetrically between magnetosheath plasmas, namely solar wind plasmas compressed by Earth's bow shock, and magnetospheric plasmas. 
The magnetosheath side has a typical magnetic field strength $\sim 20$ nT, density $\sim$ 15 cm$^{-1}$ and plasma-$\beta \sim 2$; The magnetosphere side has magnetic field strength $\sim 60$ nT, density $\sim$ 0.5 cm$^{-1}$ and plasma-$\beta \sim 0.1$ (e.g., \citet{phan96a}). 
The magnetic field shear can be an arbitrary angle $\phi$. 
Considering a planar current sheet, the x-line could develop at any angle from 0 to $\phi$, where the fields in the plane normal to this orientation have opposite signs, as suggested by \citet{cowley76a}. It is unclear if there is a simple principle to determine the orientation of the x-line in a three-dimensional (3D) system.

The first attempt to address this fundamental problem was by \citet{sonnerup74a} (also independently by \citet{gonzalez74a}), who suggested that reconnection will occur in a plane where the guide field is uniform. 
Motivated by \citet{cowley76a}, the angle that bisects the total shear has been employed in global modeling \citep{moore02a,borovsky08a,sibeck09a}.
More recently, other sophisticated models based on maximizing various physics quantities were proposed. 
\citet{swisdak07a} suggested the plane in which the reconnection outflow jets have a maximum speed. \citet{schreier10a} pointed out another possibility by maximizing the reconnection electric field (equivalent to the reconnection rate), where the formulation in \citet{cassak07b} for asymmetric reconnection could be used. Based on 2D simulations at different oblique reconnection planes, \citet{hesse13a} further proposed that the x-line orientation should be determined by maximizing the peak reconnection electric field, which was found to be proportional to the product of available magnetic energy density at both sides. The maximum of the peak reconnection electric field was shown to bisect the total magnetic shear angle $\phi$.

The principle that determines the orientation of the x-line in this simple planar current sheet could potentially guide us to find the location of reconnection in a more realistic magnetopause geometry. Global magnetospheric MHD simulations were recently performed \citep{komar15a} to compare these models, along with other ideas that predict the locations first in the global geometry with the ``orientation'' being the resulting locus that connects these locations. These predictions include maximizing the total magnetic shear angle \citep{trattner07a}, the total current density \citep{alexeev98a} and the divergence of the Poynting flux \citep{papadopoulos99a}. Another scenario suggested the x-line to be the magnetic separator simply resulting from the vacuum superposition of Earth's dipolar and solar wind magnetic fields \citep{cowley73a,siscoe01a,dorelli07a}, which was also tested \citep{komar13a}.

Observationally, the location and orientation of x-lines have been inferred from patterns of accelerated flows \citep{dunlop11b, phan06b, pu07a, scurry94a}, and patterns of precipitating ion dispersions \citep{trattner07a} during quasi-steady reconnections. Statistical studies of the flux transfer events (FTEs) generated by bursty reconnection \citep{fear12a, dunlop11a, wild07a, kawano05a}, and the global distribution of streaming energetic ion anisotropies \citep{daly84a} also provide clues. In addition, methods for locally reconstructing the reconnection geometry \citep{teh08a, shi05a, denton12a}  were developed, some of these methods \citep{shi05a, denton12a} could potentially take advantage of satellite clusters that are deployed closely, such as NASA's Magnetospheric Multiscale Mission (MMS) \citep{burch09a}.  These {\it in-situ} observations constrain theoretical modeling, however, it has been difficult to use them to distinguish between the detailed predictions of these models. To accurately determine the x-line orientation in observation is still a challenging and active research area in space study. 

In this paper, we use 3D simulations to study the orientation of x-lines in a given asymmetric planar geometry. In a similar work, \citet{schreier10a} used 3D Hall-MHD simulations, where the x-line orientation is concluded to be consistent with that of the maximized reconnection outflow speed \citep{swisdak07a} or reconnection electric field (rate) \citep{cassak07b}.  
However, a study using 3D fully kinetic simulations does not exist, where kinetic effects such as the particle streaming could be potentially important.

Furthermore, we develop a way to clearly test the orientation of a single x-line in 3D systems that develops in a controlled fashion. For reconnections that develops from a long current sheet without a perturbation, or with a perturbation similar to that in the GEM-challenge \citep{birn01a}, the x-line orientation may be strongly affected and even selected by oblique flux ropes arising from tearing instabilities in the linear phase (e.g.,  \citet{yhliu13a}). 
To avoid this, we use a spatially localized perturbation to induce a single x-line. 
This localized perturbation prevents the linear tearing instability before the development of a single x-line, and does not pre-select the orientation of x-line. The single x-line that develops with sufficient freedom appears to bisect the total magnetic shear angle for larger mass ratios. This is consistent with that suggested in \citet{hesse13a}.

The layout of this paper is the following. Section 2 describes the setup of our particle-in-cell simulations. Section 3 shows the measurements of x-line orientations in 3D simulations with $m_i/m_e=25$ and $m_i/m_e=1$ plasmas. Section 4 compares the results with 2D simulations, and the mass ratio dependency of x-line orientation is investigated. Section 5 contains the summary and discussions.

\section{Simulation setup} 

The asymmetric configuration employed \citep{hesse13a, aunai13b, pritchett08a} has the magnetic profile, ${\bf B}=B_0(0.5+S)\hat{\bf x}_0+B_{y0}\hat{\bf y}_0$ where $S=\alpha_1\mbox{tanh}(z/\lambda)$.  This corresponds to a shear angle $\phi=180^\circ-\mbox{tan}^{-1}[(B_{y0}/B_0)/(0.5+\alpha_1)]-\mbox{tan}^{-1}[(B_{y0}/B_0)/(0.5-\alpha_1)]$ across the sheet. The plasma has density $n=n_0[1-\alpha_2(S+S^2)/3]$ and an uniform total temperate $T=3B_0^2/(8\pi n_0 \alpha_2)$. We choose $\alpha_1=\alpha_2=1$, then the resulting $B_{2x0}=1.5B_0$, $B_{1x0}=0.5B_0$ and $n_2=n_0/3$, $n_1=n_0$. Here the subscripts ``1'' and ``2'' indicate the magnetosphere and the magnetosheath sides respectively. We use a uniform guide field $B_{y0}=B_0$ then the total magnetic shear agnle $\phi\sim 82.87^\circ$. The temperature ratio is $T_i/T_e=5$, and the ratio of electron plasma to gyro-frequency is $\omega_{pe}/\Omega_{ce}=4$. Here, $\omega_{pe}\equiv(4\pi n_0 e^2/m_e)^{1/2}$ and $\Omega_{ce}\equiv eB_0/m_e c$.

In this paper, fully kinetic simulations were performed using the particle-in-cell code -{\it VPIC} \citep{bowers09a}. Densities are normalized by density $n_0$, time is normalized by the ion gyro-freqency $\Omega_{ci}$, velocities are normalized by Alfv\'enic speed $V_A\equiv B_0/(4\pi n_0 m_i)^{1/2}$, and spatial scales are normalized by the inertia length $d_j\equiv c/\omega_{pj}$, where $j=e, i$ for electron or ion respectively. 

For the rest of this paper, the x-line orientation will be quantified using the angle $\theta$ respect to the ${\bf y}_0$-axis.
The simulation box can be rotated to $\hat{\bf x}=\mbox{cos}\theta_{box}\hat{\bf x}_0+\mbox{sin}\theta_{box}\hat{\bf y}_0$ and $\hat{\bf y}=\mbox{sin}\theta_{box}\hat{\bf x}_0-\mbox{cos}\theta_{box}\hat{\bf y}_0$. In a 2D system, this machinery allows us study the reconnection with a pre-selected x-line orientation $\theta=\theta_{box}$. The in-plane magnetic field vanishes at $z_{n}=\lambda \mbox {tanh}^{-1}\{-[0.5+(B_{y0}/B_0)\mbox{tan}\theta_{box}]/\alpha_1\}$.

The primary 3D run (case {\bf k} in Table 1) discussed in detail uses $m_i/m_e=25$ and has a domain size of $L_x \times L_y \times L_z=64d_i \times 64d_i \times 16d_i$ with $1024 \times 512 \times 256$ cells. The simulation domain is rotated to $\theta=+10^\circ$, this does not affect the conclusions in this paper. The boundary conditions are periodic both in the x- and y-directions, while in the z-direction are conducting for fields and reflecting for particles. We use $150$ particles per cell. The half-thickness of the initial sheet is $\lambda= 0.8 d_i$. In addition to 3D simulations, 2D runs with $m_i/m_e=1, 4, 25, 100$ and $256$ are also conducted to study the mass ratio dependency. These runs are listed in Table 1.

\begin{figure}[!htb]
  \centering
    {\includegraphics[trim = 0mm 0mm 0mm 0mm, clip, width=55mm]{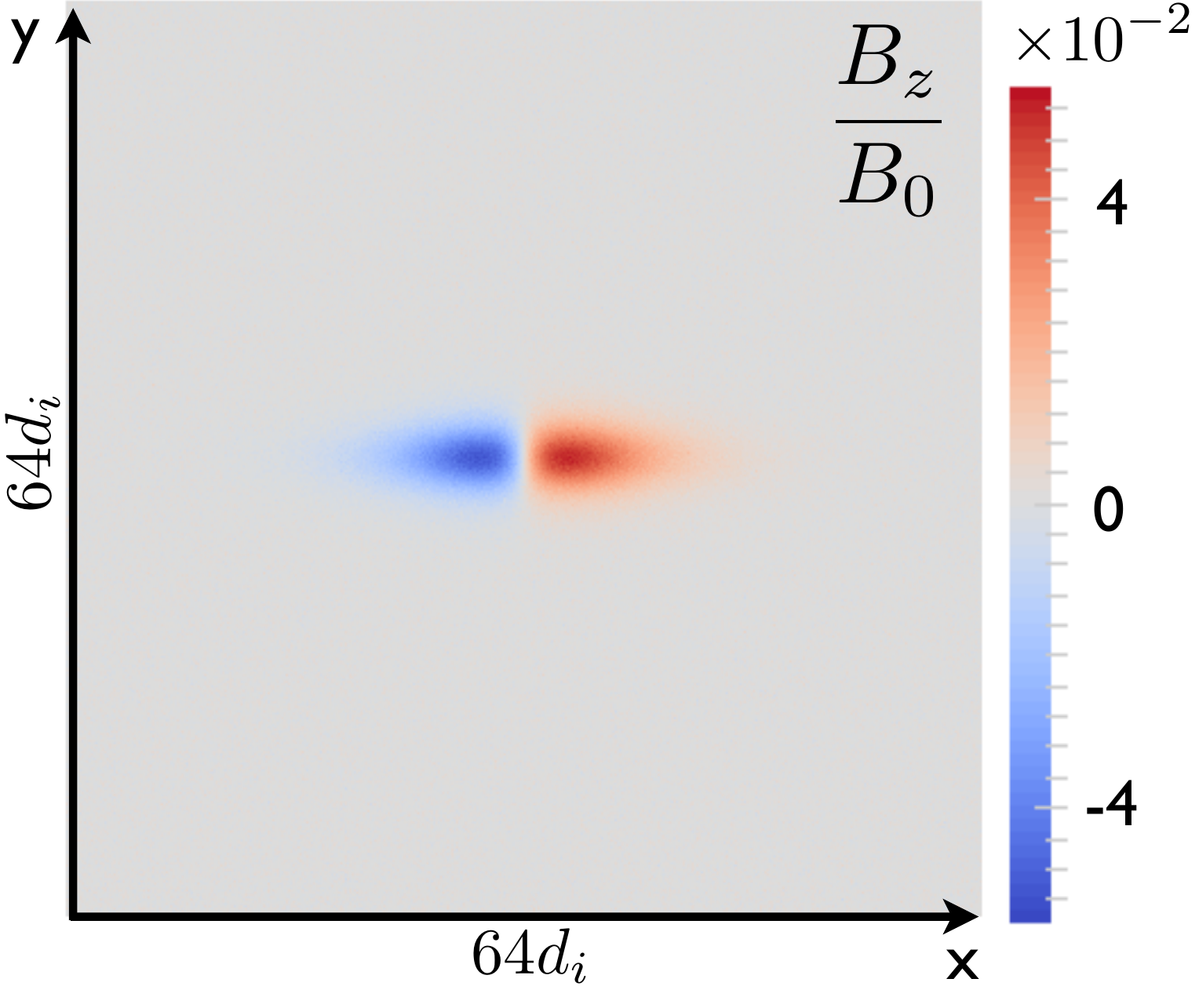}}
    \caption{$m_i/m_e=25$. The localized perturbation with amplitude $\delta B_z =0.05 B_0$ initially imposed in the simulation box with the y-direction aligned to $\theta=10^\circ$. This 2D plane is taken at $z=z_p=-0.5493 \lambda$.}
    \label{pert}
\end{figure}

To study the simplest situation with a single x-line, we want to avoid the development of tearing instabilities before a well-defined x-line forms. We use a perturbation localized in the x-direction since the tearing mode is more stable in a short current sheet. In addition, a perturbation being uniform in the y-direction might pre-select the orientation. Therefore, we further localize the perturbation in the y-direction so that the single x-line can develop with sufficient freedom. The perturbation used in $m_i/m_e=25$ cases is illustrated in Fig.~\ref{pert}.
The perturbation has the functional form $\tilde{B}_z\propto \mbox{cos}[\pi (z-z_p)/L_z]\times\mbox{sin}(2\pi x/L_x) \mbox{exp}(-|x|/L_{p1}) \times f(y)$, where $f(y)=\mbox{tanh}[(y+L_{p2})/L_{p3}]-\mbox{tanh}[(y-L_{p2})/L_{p3}]$. $\tilde{B}_x$ is derived using $\nabla\cdot \tilde{\bf B}=0$ and $\tilde{B}_y=0$. The peak value of the perturbation is $\delta B_z=0.05 B_0$. For runs with mass ratio $m_i/m_e=1$, $4$, $25$ and $256$, we choose $L_{p1}=L_x/40$, $L_x/8$, $L_x/20$ and $L_x/20$; $L_{p2}/d_i=1$, $0.5$, $1$ and $1$; $L_{p3}/d_i=2$, $3$, $2$ and $2$ respectively. To simplify the comparison, we fix $z_p=-0.5493 L$ for all cases. This is the location where the in-plane magnetic vanishes at the $\theta=0^\circ$ plane. 

It may be argued that the orientation of these single x-lines will be constrained to meet the resonant condition imposed by periodic boundaries, i.e., $(L_y/L_x)\mbox{tan}(\theta+\theta_{box})$ being a rational number, which is equivalent to the safety factor in fusion Tokamaks (e.g., \citet{beidler11a}). An extensive study on the effect of periodic boundary using mass ratio $m_i/m_e=4$ is performed, but not shown here. We find that the x-line in a large enough simulation box develops into the same orientation even with different box aspect ratio and box orientation. We conclude that the localization of x-line as described mitigates this effect from periodic boundaries.

\section{3D simulation results}

\begin{figure}[!htb]
  \centering
    {\includegraphics[trim = 0mm 0mm 0mm 0mm, clip, width=100mm]{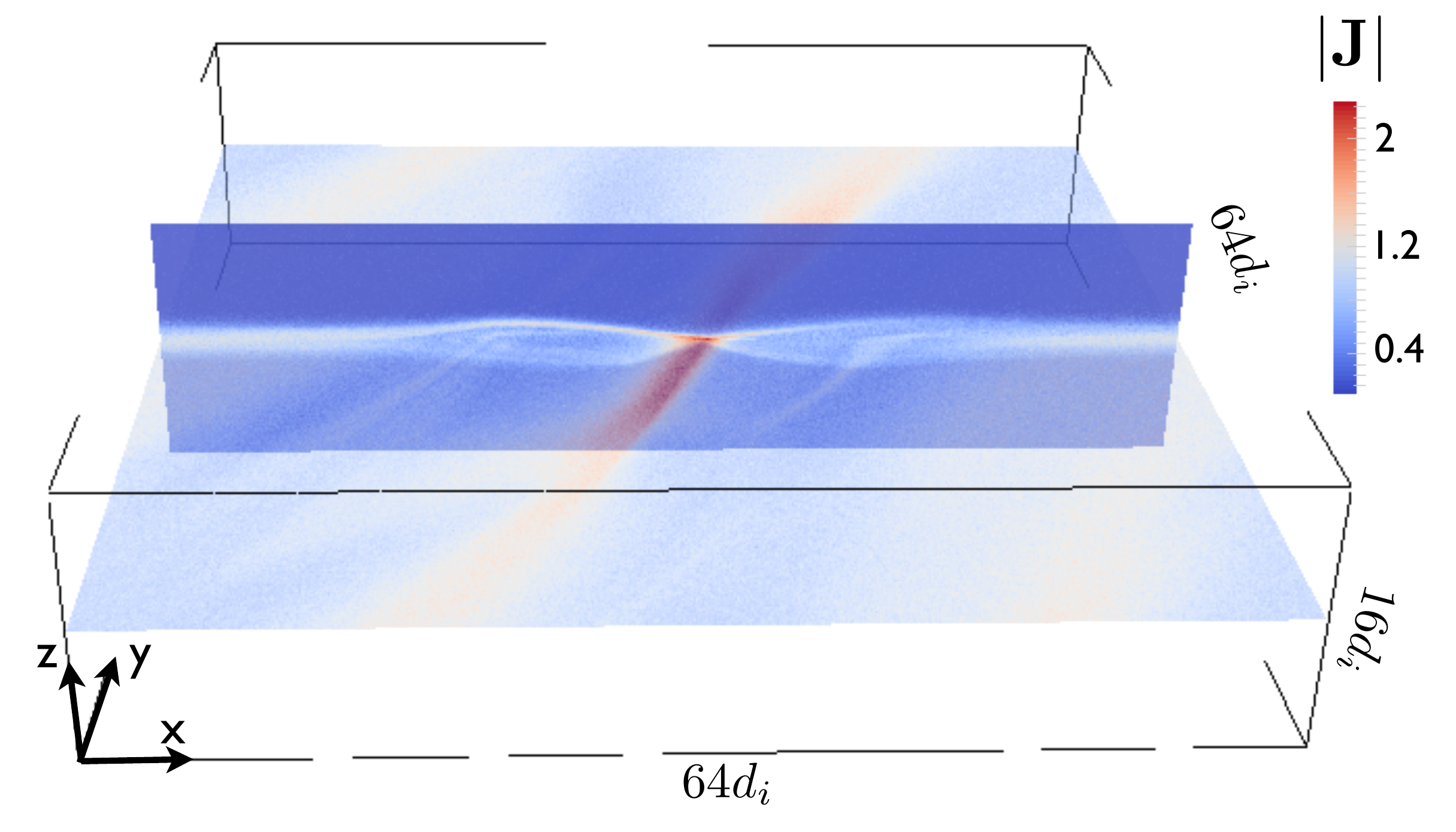}}
    \caption{Case-${\bf k}$ with $m_i/m_e=25$ at $60/\Omega_{ci}$. The global structure of the x-line is shown as $|J|$ in 2D planes at $y=0$ and $z=0$. The y-direction of the simulation box is aligned to $\theta=10^\circ$.}
    \label{local_mime25_10d_3D}
\end{figure}

\begin{figure}[!htb]
  \centering
    {\includegraphics[trim = 0mm 0mm 0mm 0mm, clip, width=100mm]{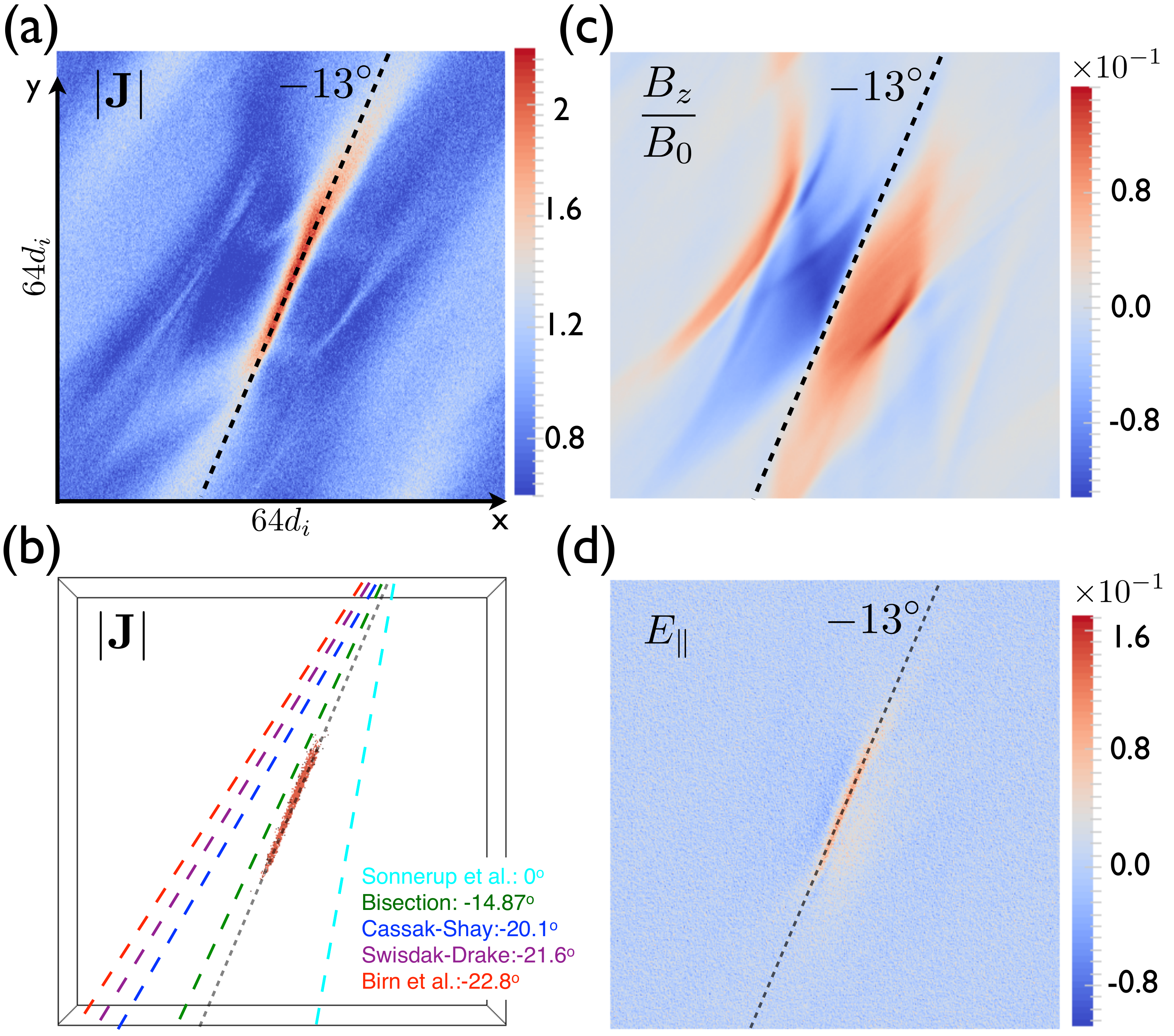}}
    \caption{Case-${\bf k}$ with $m_i/m_e=25$ at $60/\Omega_{ci}$. In (a), the current density $|J|$ at $z=0$.  In (b), the 3D iso-surface of $|J|=2$ is overlaid with theoretical predictions. In (c), the reconnected magnetic field component $B_z$ at $z=0$. In (d), the parallel (non-ideal) electric field $E_\|$ normalized to $V_A B_0/c$ is displayed at $z=0$. Black dotted lines have $\theta=-13^\circ$.}
    \label{local_mime25_10d}
\end{figure}

With the localized perturbation, a well-defined x-line emerges near the center of the simulation box. Figure \ref{local_mime25_10d_3D} shows the total current density $|{\bf J}|$ of the primary case at time $60/\Omega_{ci}$, when a single x-line with large scale outflows has developed. The planes are cuts at $z=0$ and $y=0$. In order to measure the orientation of the x-line, we focus on the $x-y$ plane (top-view) in Fig.~\ref{local_mime25_10d}. The current density is shown in panel (a) and a black-dotted line of $\theta=-13^\circ$ is overlaid for comparison. To avoid a potential dependency on the choice of the $x-y$ plane, the 3D iso-surface of $|{\bf J}|=2$ is plotted in Fig.~\ref{local_mime25_10d}(b), which further justifies the measurement of this angle. The reconnected magnetic field $B_z$ is shown in Fig.~\ref{local_mime25_10d}(c). The region with $B_z=0$ indicates the topological separator and follows the same black-dotted line. Note that the average magnitude of $B_z$ is around $\sim O(0.1 B_0)$, as expected in the nonlinear stage of reconnection. Figure.~\ref{local_mime25_10d}(d) depicts the non-ideal electric field $E_\|$, which traces the diffusion region of magnetic reconnection, also shows the same orientation. 
Fig. \ref{local_mime25_10d}(b) and (d) suggest the x-line extension $\approx 20 d_i=100 d_e$, and interestingly the x-line does not appear to extend much longer at later time. A similar finite extension was observed in symmetric reconnection simulations \citep{shay03a}.

\begin{figure}[!htb]
  \centering
    {\includegraphics[trim = 0mm 0mm 0mm 0mm, clip, width=100mm]{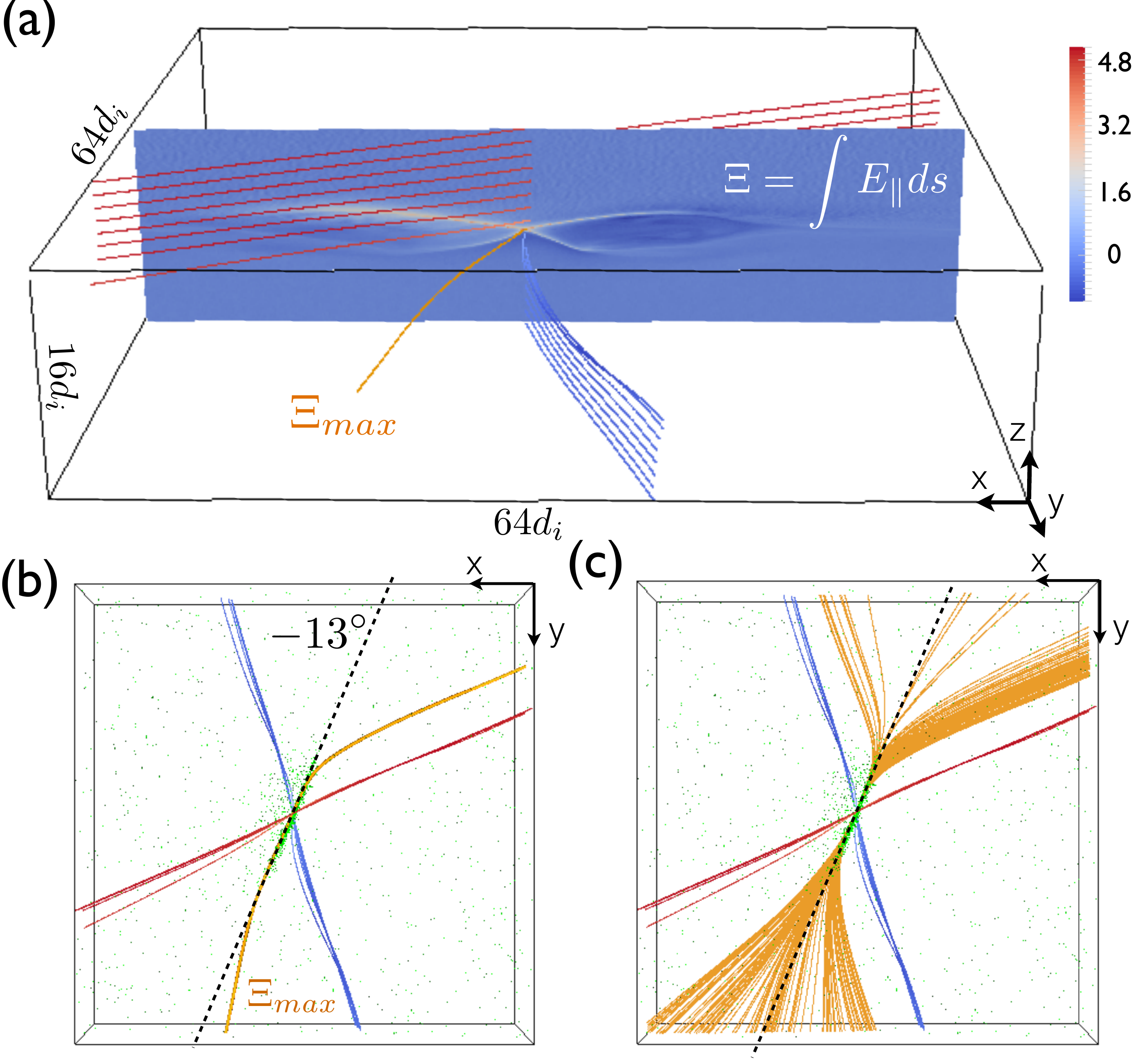}}
    \caption{Case-${\bf k}$ with $m_i/m_e=25$ at $60/\Omega_{ci}$. In (a), the 2D map of quasi-potential $\Xi=\int{E_\| ds}$ at $y=0$. The field line with $\Xi_{max}$ is colored in yellow and sample field lines distributed vertically are colored in red ($B_x > 0$) and blue ($B_x < 0$). Panel (b) is the top view of (a) with the iso-surface of $E_\|=0.08 V_AB_0/c$ overlaid in green. Panel (c) has a similar format of (b), but depicts 100 magnetic field lines (yellow) traced from seeds that are evenly distributed inside a sphere centered at the location of $\Xi_{max}$ with radius $0.1d_e$.}
    \label{potential}
\end{figure}

To evaluate the global reconnection rate, we apply the general magnetic reconnection theory (GMR) \citep{schindler88a, hesse88a, hesse93a} on this three-dimensionally localized x-line. GMR theory points out the importance of evaluating the integration of the parallel electric field $E_\|$ along magnetic field lines, $\Xi\equiv\int{E_\| ds}$, especially for field lines that thread the ideal region ($E_\|=0$) through the non-ideal region ($E_\| \neq 0$) to the ideal region at another end. The maximum value, $\Xi_{max}\equiv \mbox{max}[\Xi(x,z)]$, is the global reconnection rate. This will be an accurate measure of reconnection rate since the net contribution of electrostatic component in $E_\|$, that is not directly relevant to reconnection, will vanish in this integration. The integration reduces the 3D system to a 2D map of $\Xi$, as shown in Fig.~\ref{potential}(a). This $\Xi$ map in the $y=0$ plane is generated by integrating $E_\|$ along field lines for $30 d_i$ arc-length at both sides of the $y=0$ plane. 
We can then identify the location of $\Xi_{max}$ on this 2D map and trace the magnetic field line from this seed point (yellow). This magnetic field line that carries $\Xi_{max}$ is expected to be tangential to the x-line locally around the diffusion region if the diffusion region is quasi-2D for a reasonably long extension. 
For comparison, we also trace 15 field lines seeded evenly along the z-direction at the same x and y coordinate of $\Xi_{max}$. These sample field lines with positive (negative) $B_x$ are colored in red (blue).

Figure \ref{potential}(b) shows the top-view of these field lines overlaid with the iso-surface of $E_\|=0.08V_AB_0/c$ (green). 
The field line with $\Xi_{max}$ (yellow) appears to pass through the non-ideal region and is tangential to the black-dashed line with orientation $\theta=-13^\circ$. This orientation approximately bisects the total magnetic shear angle across the current sheet (i.e., the angle between the red and blue field lines). It may be argued that the field line behavior may be sensitive to the choice of seed points due to the chaotic nature of magnetic field lines \citep{boozer12a}. Hence, to get a more conclusive measurement, we also seed 100 points evenly distributed inside a sphere of radius $0.1d_e$ centered at the location of $\Xi_{max}$. These field lines traced from these seeds are shown in Fig.~\ref{potential}(c) in yellow. They align with orientation $\theta\approx-13^\circ$ inside the non-ideal region (green), then separate quickly from each other outside the non-ideal region. The global reconnection rate is $\Xi_{max} \approx 4.8 V_AB_0d_e/c$. Divided by the length of the x-line $\approx 100 d_e$, the spatially-averaged 2D rate is roughly $0.048V_AB_0/c$, quantitatively similar to the peak $E_\|$ measured in the corresponding 2D simulation at this orientation (Fig.~\ref{2D_rates}(c)). This further justifies that this 3D x-line is at its nonlinear phase. 

In Fig.~\ref{local_mime25_10d}(b), theoretical predictions of x-line orientation \citep{sonnerup74a, swisdak07a, cassak07b, schreier10a, birn10a, hesse13a} are plotted as dashed lines using different colors. Note that a prediction based on the reconnection electric field in \citet{birn10a} is also presented here, where a more accurate energy equation is considered to improve the Cassak-Shay formula \citep{cassak07b}. The ratio of specific heats $5/3$ is used. The closest prediction is the angle of bisection with $\theta=-14.87^\circ$ \citep{hesse13a, moore02a, borovsky08a, sibeck09a}. Using this same asymmetric configuration with $m_i/m_e=25$, \citet{hesse13a} found a relation between the peak reconnection electric field and the available magnetic energy for reconnection  $E_{rec}\propto B^2_{1,rec} B^2_{2,rec}$. The orientation that bisects the total magnetic shear angle maximizes this $E_{rec}$. 

\begin{figure}[!htb]
  \centering
    {\includegraphics[trim = 0mm 0mm 0mm 0mm, clip, width=100mm]{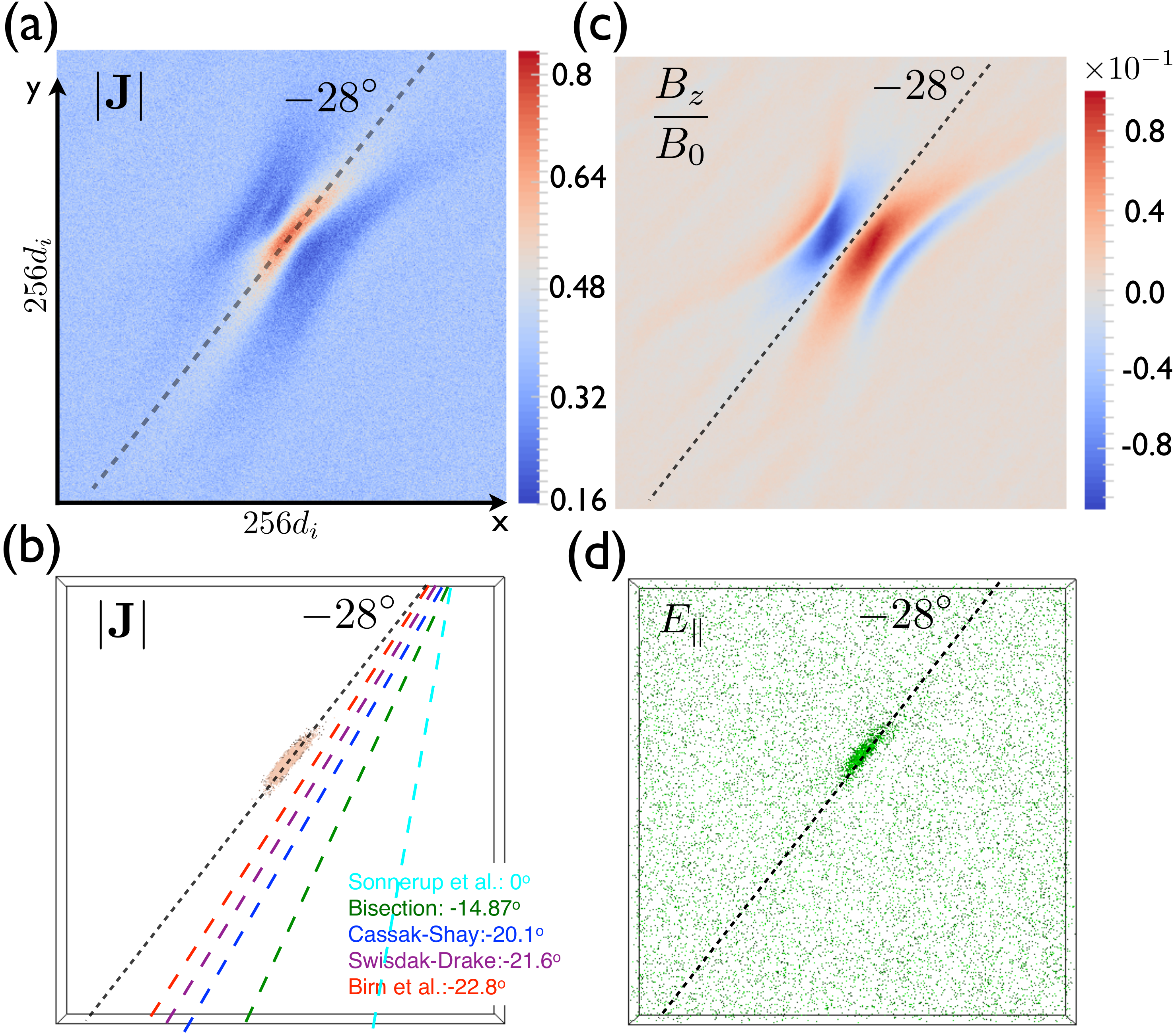}}
    \caption{Case-${\bf a}$ with $m_i/m_e=1$ at $300/\Omega_{ci}$. In (a), the current density $|J|$ at $z=-0.4d_e$.  In (b), the 3D iso-surface of $|J|= 0.62$ is overlaid with theoretical predictions. In (c), the reconnected magnetic field component $B_z$ at $z=-0.4d_e$. In (d), the iso-surface of parallel (non-ideal) electric field $E_\|=0.016V_A B_0/c$. Black dotted lines have $\theta=-28^\circ$.}
    \label{local_mime1_10d}
\end{figure}

While the agreement between this 3D simulation and the theoretical prediction is excellent, it is important to test if this bisection orientation is generic. We perform a similar 3D simulation in electron-positron plasmas with mass ratio $m_i/m_e=1$ to test the mass ratio dependency. The measurement of $|{\bf J}|$, $B_z $ and $E_\|$ displayed in Fig.~ \ref{local_mime1_10d} consistently suggest an angle $\approx -28^\circ$, which is larger than the bisection angle. However, all existing analytical predictions \citep{sonnerup74a, swisdak07a, cassak07b, schreier10a, birn10a, hesse13a} remain the same as indicated in Fig.~\ref{local_mime1_10d}(b), since they do not have the mass ratio dependency. This is investigated further in the following section.

\section{2D modeling and predictions} 

\begin{figure}[!htb]
  \centering
    {\includegraphics[trim = 0mm 0mm 0mm 0mm, clip, width=110mm]{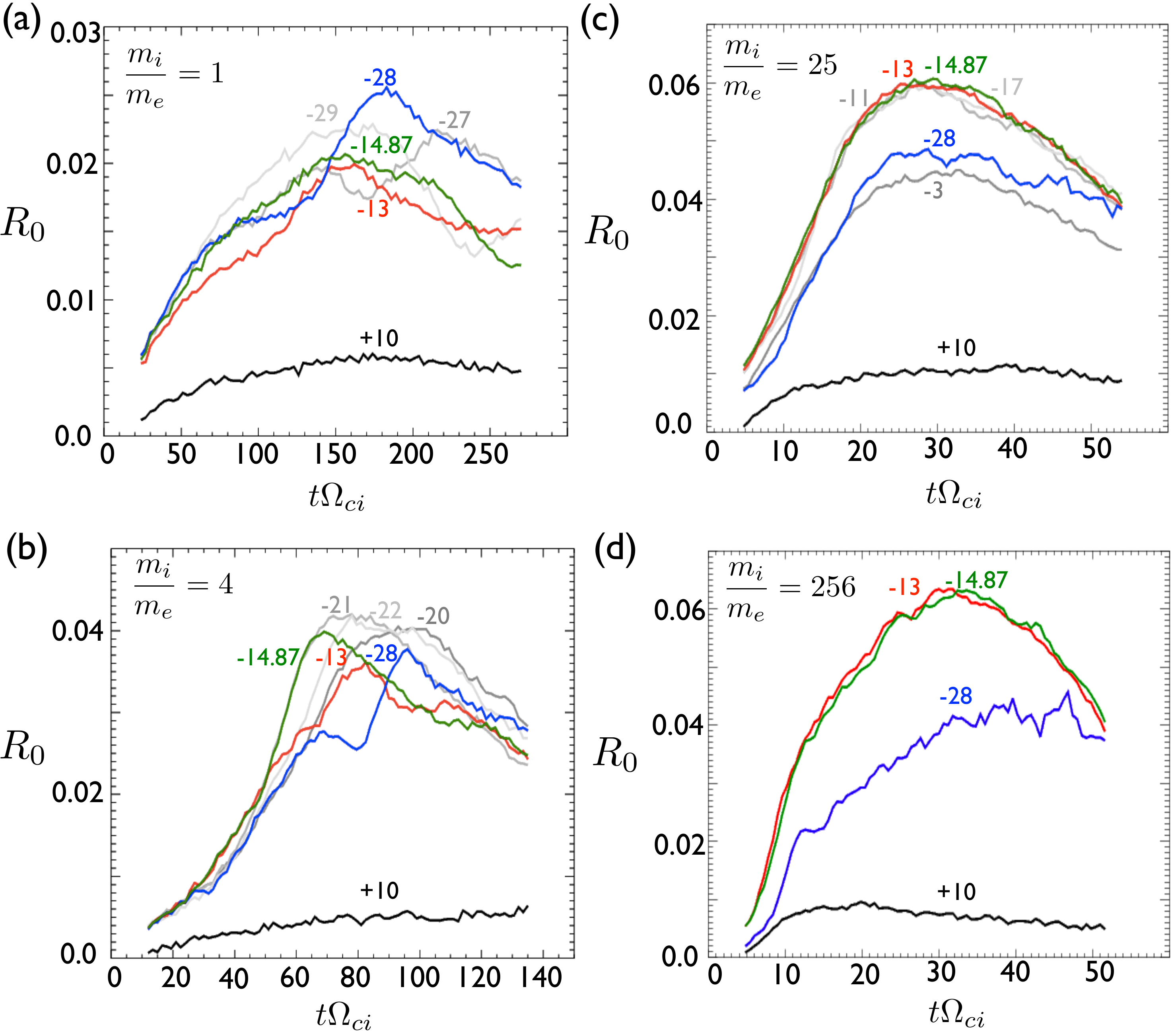}}
    \caption{Reconnection rate $R_0$ at sample orientations with mass ratios (a) $m_i/m_e=1$, (b) $m_i/m_e=4$, (c) $m_i/m_e=25$ and (d) $m_i/m_e=256$.}
    \label{2D_rates}
\end{figure}

To understand the difference with a lower mass ratio, we go back to 2D simulations at oblique planes (i.e., $\theta_{box}\neq 0$). Unlike 3D, the advantage of using 2D simulation is that we can choose the orientation of the x-line, which is out of the 2D plane. We study the evolution of the reconnection rate $R_0\equiv \left<\partial_t \psi/\partial t\right> /(V_A B_0)$ at different orientations in Fig.~\ref{2D_rates}. Here $\psi$ is the difference of the flux function $A_y$ between the primary x-  and o-points, which are the saddle point and local maximum of $A_y$ respectively.
As shown in Fig.~\ref{2D_rates}(c), with $m_i/m_e=25$ the orientation that maximizes the peak reconnection rate are consistent with the bisection angle with $\theta=-14.87^\circ$ \citep{hesse13a}.  However, with a lower mass ratio the orientation that maximizes the peak rate shifts to a larger angle as shown in Fig.~\ref{2D_rates}(a)-(b).  The suggested angle from these 2D simulations with electron-positron plasmas ($m_i/m_e=1$) is $\theta \approx -28^\circ$, consistent with the orientation measured in the 3D system (Fig.~\ref{local_mime1_10d}). This further suggests that 2D models are sufficient to capture the physics that determines the x-line orientation in 3D systems, and this orientation maximizes the peak reconnection rate among these 2D oblique planes.
With a larger mass ratio, the bisection angle persists to maximizes the peak rate in 2D simulations, as shown in Fig.~\ref{2D_rates} (d) with $m_i/m_e=256$. While a 3D simulation similar to that of Fig.~\ref{local_mime25_10d_3D} with a realistic mass ratio $m_i/m_e=1836$ is impossible with current computational capability, the consistency between 3D and 2D simulations demonstrated here suggests that the x-line may still bisect the magnetic shear with real mass ratio $m_i/m_e=1836$. This prediction is directly relevant to the reconnection events at Earth's magnetopause.

To explain the mass ratio dependency, we notice that secondary plasmoids are generated to cause fluctuations in the reconnection rates shown in Fig.~\ref{2D_rates}(a)-(b), even though we used the localized perturbation and a thicker initial sheet. In contrast, the induced singe x-line in plasmas of higher mass ratios (e.g., $m_i/m_e=25, 256$ in Fig.~\ref{2D_rates}(c)-(d)) does not generate secondary plasmoids, presumably because the Hall effect arising from the mass ratio difference prevents the opened reconnection exhaust from collapsing \citep{shay99a, stanier15a}, and hence makes tearing modes more stable \footnote{However, the reconnection rate has the same order regardless the difference in mass ratio.}.
This further motivates us to conjecture that the physics of tearing instability may play some role, that is more apparent with a lower mass ratio. The tearing instability is driven by the filamentation tendency of current sheet. In principle, the nonlinear current sheet of the single x-line could still be subject to the same filamentation tendency. 

To investigate this, we use a current sheet of $d_e$-scale thick. The half-thickness $\lambda=1.36d_e$ is taken to be the mean value of the inertial lengths at both sides (i.e., $d_e$ and $\sqrt{3}d_e$). This thickness mimics the current sheet scale observed in the nonlinear stage of reconnection. In Fig.~\ref{tearing} (a), we show the tearing modes that spontaneously grow in this $d_e$-scale current sheets without any perturbation. At the time of measurement, the tearing mode amplitude is still small, $\delta B_z/B_0\sim O(10^{-3})$, and hence justifies its linear stage. The dominant tearing modes, presumably the fastest growing tearing mode, has a similar orientation $\theta\approx -28^\circ$ as that of the single x-line observed in Fig.~\ref{local_mime1_10d}. This supports our conjecture on the role of the tearing instability. 
In addition, we do the same experiment with higher mass ratio $m_i/m_e=25$. Interestingly, the dominant tearing modes manifest an angle $\theta\approx -13^\circ$  as shown in Fig.~\ref{tearing}(b), that is also consistent with the x-line orientation in Fig~\ref{local_mime25_10d}. These results imply that tearing modes may have some relation to the peak reconnection rate measured in 2D simulations (Fig.~\ref{2D_rates}), and hence the bisection solution in the large mass ratio limit.

\section{Summary and Discussion} 

We demonstrate that in the large mass ratio limit the x-line bisects the total magnetic shear angle across the current sheet, at least in the 3D simulations presented here.  The orientation can generally be predicted by scanning through a series of 2D simulations to find the orientation that maximizes the peak reconnection electric field.
This result serves as a practical prediction to reconnection events at Earth's magnetopause.

\begin{figure}[!htb]
  \centering
    {\includegraphics[trim = 0mm 0mm 0mm 0mm, clip, width=100mm]{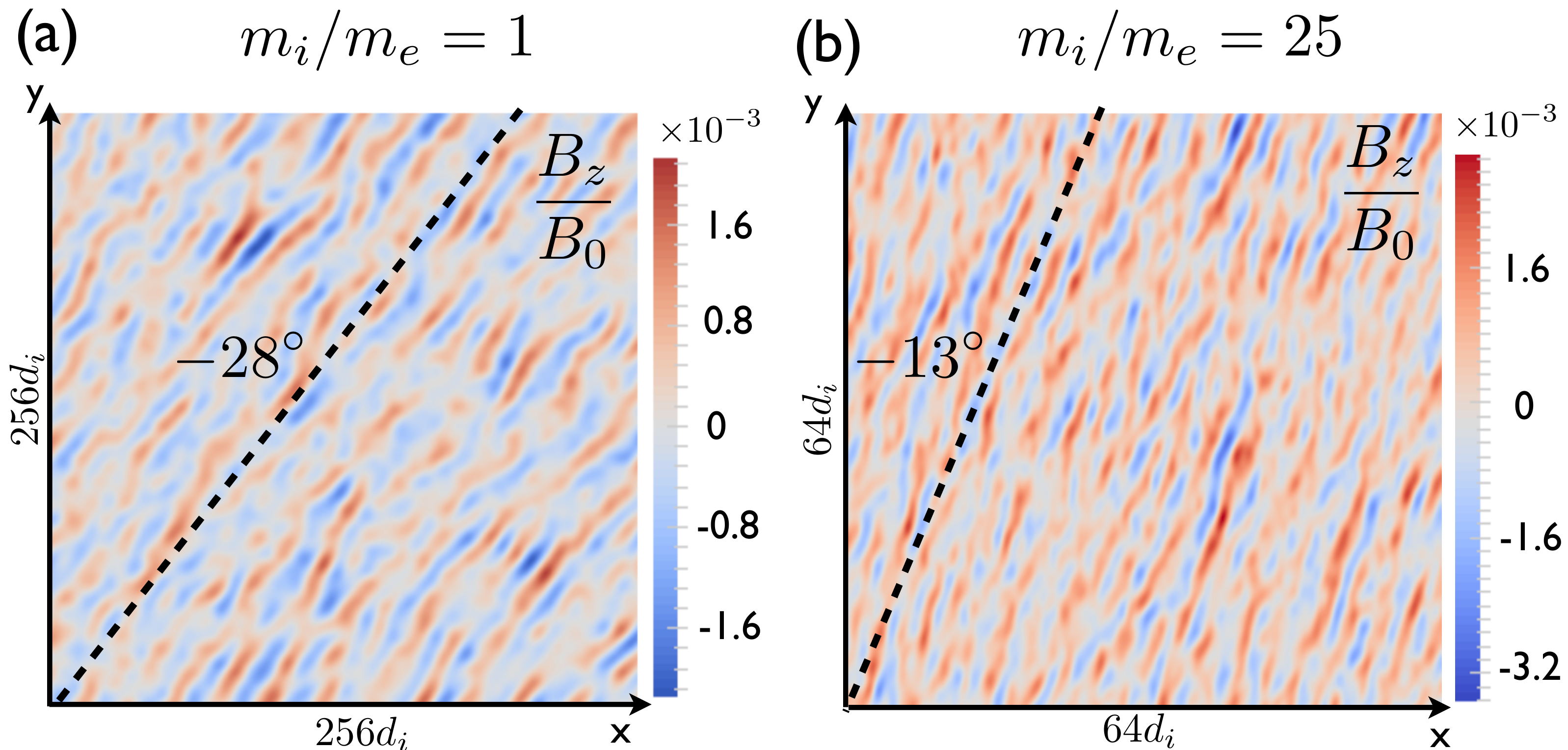}}
    \caption{Tearing modes in a electron-scale current sheet of half-initial thickness $\lambda=1.36d_e$.  In (a), case-${\bf d}$ with $m_i/m_e= 1$ at time $51.25/\Omega_{ci}$. $B_z$ is time-averaged for $12.5/\Omega_{ci}$. The black dotted line has $\theta=-28^{\circ}$. In (b), case-${\bf n}$ with $m_i/m_e= 25$  at time $3.5/\Omega_{ci}$. $B_z$ is time-averaged for $2/\Omega_{ci}$. The black dotted line has $\theta=-13^{\circ}$.}
    \label{tearing}
\end{figure}

The fact that $d_e$-scale tearing modes share the same orientation as a nonlinear single x-line may have profound implications. 
The tearing instability is driven by the filamentation tendency of the current sheet. In principle, the nonlinear current sheet of the single x-line could still be subject to the same tendency, and consequently develop into a state that is marginally stable to the tearing instability. The linear tearing mode hence may provide predictions on some properties of the x-line, for instance, the orientation shown here and maybe the spatial scale of the x-line \citep{yhliu14a}. 
To study these $d_e$-scale tearing modes in 3D simulations with a larger mass ratio is still computationally feasible, since we only need to check the linear phase and the tearing instability's growth rate is large for a $d_e$-scale current sheet. In terms of ion gyro-frequency, the growth rate \citep{yhliu13a,daughton11a}  is $\gamma/\Omega_{ci} \sim (d_e/\lambda)^3(m_i/m_e)(\rho_e/d_e)$. In this Harris-type equilibrium $\beta\sim O(1)$, then $\rho_e/d_e \sim O(1)$, hence the growth rate is proportional to the mass ratio $m_i/m_e$ in a $d_e$-scale current sheet. Therefore, simulations like Fig.~\ref{tearing} could also serve as a useful indicator in predicting the x-line orientation. 

\begin{figure}[!htb]
  \centering
    {\includegraphics[trim = 0mm 0mm 0mm 0mm, clip, width=90mm]{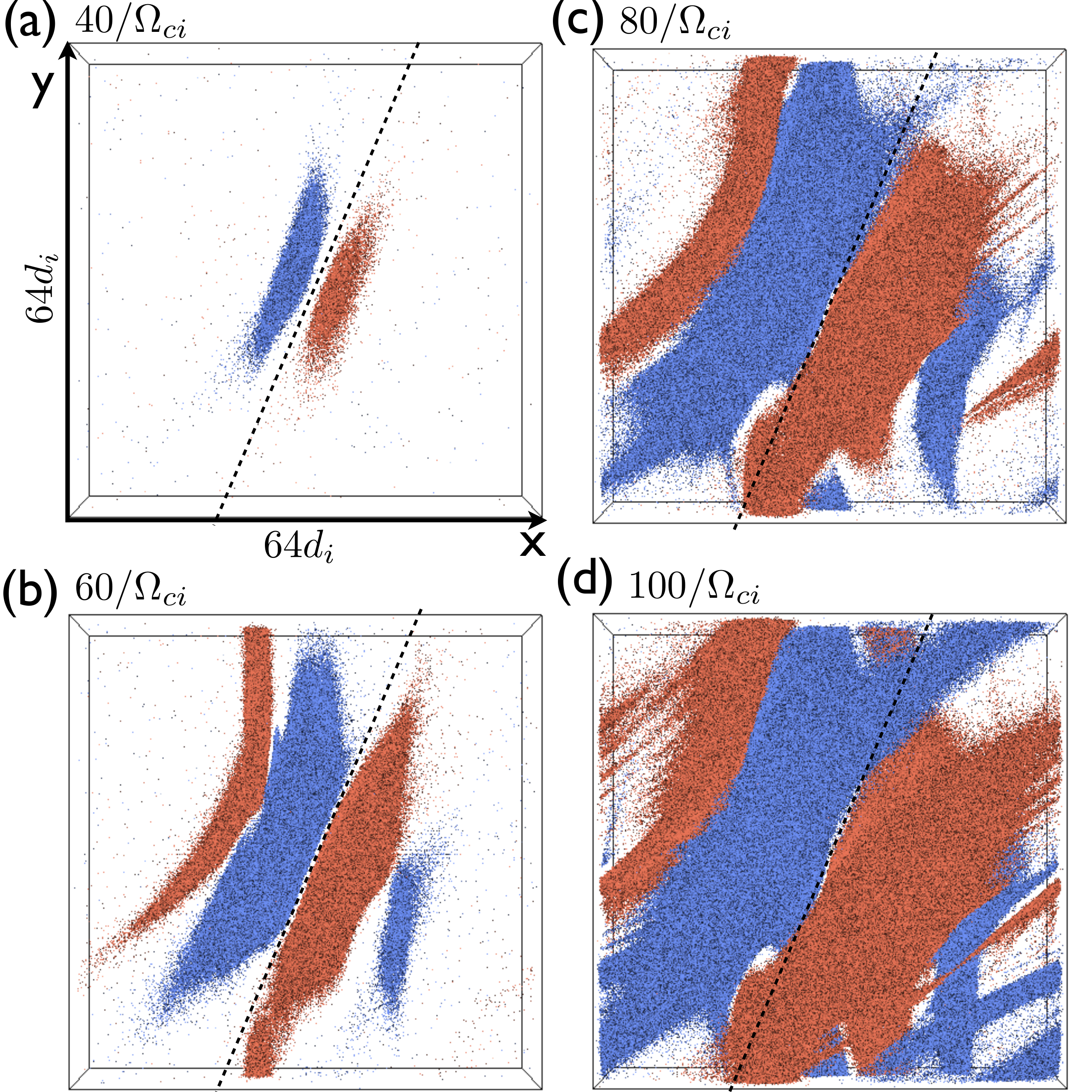}}
    \caption{Case-${\bf k}$ with $m_i/m_e=25$. The iso-surface of the reconnected magnetic field component $B_z/B_0=0.08$ is shown for time (a) $t=40/\Omega_{ci}$, (b) $t=60/\Omega_{ci}$, (c) $t=80/\Omega_{ci}$ and (d) $t=100/\Omega_{ci}$. The black dotted lines have $\theta=-13^\circ$.}
    \label{Bz_evolution}
\end{figure}

Some caveats and limitations need to be kept in mind.
First, at late times, periodic boundaries may start to affect and secondary flux ropes (i.e., 3D version of plasmoids) develop along the separatrix \citep{daughton11a} of these single x-lines. These oblique flux ropes intertwine with each other and complicate the current sheet as seen in Fig.~\ref{Bz_evolution}(d), where a definite measurement of the x-line orientation becomes difficult. However, the orientation of the primary topological separator in Fig.~\ref{Bz_evolution}(d) appears to remain similar. Second, for multiple x-lines that develop from periodic tearing modes in a current sheet without a localized perturbation, the orientation could be strongly affected by the nonlinear flux-ropes \citep{yhliu13a}. Third, this study employs one possible asymmetric configuration where the stabilization by the diamagnetic drift \citep{swisdak10a, phan10a} is weak. The difference of plasma-$\beta$ between both sides is $\Delta\beta \sim 2(\delta/d_i)\mbox{tan}(\phi/2) \sim 2$ if the current sheet thickness $\delta \sim d_i$ is assumed. Future work will explore the regime with $\Delta\beta \gg  2(\delta/d_i)\mbox{tan}(\phi/2)$ to demonstrate the effects of diamagnetic drifts on the development of x-lines in 3D systems.

\begin{acknowledgments}
Y. -H. Liu thanks for helpful discussions with W. Daughton, D. G. Sibeck, C. M. Komar, N. Bessho, J. C. Dorelli, P. Cassak, N. Aunai, L. -J Chen, D. Wendel, M. L. Adrian, I. Honkonen and L. B. Wilson III. We are grateful for support from NASA through the NASA Postdoctoral Program and MMS mission. Simulations were performed with LANL institutional computing and NASA Advanced Supercomputing.  
\end{acknowledgments}

\begin{thebibliography}{51}
\providecommand{\natexlab}[1]{#1}
\expandafter\ifx\csname urlstyle\endcsname\relax
  \providecommand{\doi}[1]{doi:\discretionary{}{}{}#1}\else
  \providecommand{\doi}{doi:\discretionary{}{}{}\begingroup
  \urlstyle{rm}\Url}\fi

\bibitem[{\textit{Alexeev et~al.}(1998)\textit{Alexeev, Sibeck, and
  Bobrovnikov}}]{alexeev98a}
Alexeev, I.~I., D.~G. Sibeck, and S.~Y. Bobrovnikov (1998), Concerning the
  location of magnetopause merging as a function of the magnetopause current
  sheet, \textit{J. Geophs. Res}, \textit{103}(A4), 6675--6684.

\bibitem[{\textit{Aunai et~al.}(2013)\textit{Aunai, Hesse, Zenitani,
  Kuznetsova, Black, Evans, and Smets}}]{aunai13b}
Aunai, N., M.~Hesse, S.~Zenitani, M.~Kuznetsova, C.~Black, R.~Evans, and
  R.~Smets (2013), Comparison between hybrid and fully kinetic models of
  asymmetric magnetic reconnection: Coplanar and guide field configuration,
  \textit{Phys. Plasmas}, \textit{20}, 022,902.

\bibitem[{\textit{Beidler and Cassak}(2011)}]{beidler11a}
Beidler, M.~T., and P.~A. Cassak (2011), Model for incomplete reconnection in
  sawtooth crashes, \textit{Phys. Rev. Lett.}, \textit{107}, 255,002.

\bibitem[{\textit{Birn et~al.}(2001)}]{birn01a}
Birn, J., et~al. (2001), Geospace {E}nvironmental {M}odeling ({GEM}) magnetic
  reconnection challenge, \textit{J. Geophys. Res.}, \textit{106}(A3),
  3715--3719.

\bibitem[{\textit{Birn et~al.}(2010)\textit{Birn, Borovsky, Hesse, and
  Schindler}}]{birn10a}
Birn, J., J.~E. Borovsky, M.~Hesse, and K.~Schindler (2010), Scaling of
  asymmetric reconnection in compressible plasmas, \textit{Phys. Plasmas},
  \textit{17}, 052,108.

\bibitem[{\textit{Boozer}(2012)}]{boozer12a}
Boozer, A.~H. (2012), Separation of magneitc field lines, \textit{Phys.
  Plamas}, \textit{19}, 112,901.

\bibitem[{\textit{Borovsky}(2008)}]{borovsky08a}
Borovsky, J.~E. (2008), The rudiments of a theory of solar wind/magnetosphere
  coupling derived from first principle, \textit{J. Geophs. Res}, \textit{113},
  A08,228.

\bibitem[{\textit{Bowers et~al.}(2009)\textit{Bowers, Albright, Yin, Daughton,
  Roytershteyn, Bergen, and Kwan}}]{bowers09a}
Bowers, K., B.~Albright, L.~Yin, W.~Daughton, V.~Roytershteyn, B.~Bergen, and
  T.~Kwan (2009), Advances in petascale kinetic simulations with {VPIC} and
  {R}oadrunner, \textit{Journal of Physics: Conference Series}, \textit{180},
  012,055.

\bibitem[{\textit{Burch and Drake}(2009)}]{burch09a}
Burch, J.~L., and J.~F. Drake (2009), Reconnecting magnetic fields, \textit{Am.
  Sci.}, \textit{97}, 392.

\bibitem[{\textit{Cassak and Shay}(2007)}]{cassak07b}
Cassak, P.~A., and M.~A. Shay (2007), Scaling of asymmetric magnetic
  reconnection: General theory and collisional simulations, \textit{Phys.
  Plasmas}, \textit{14}, 102,114.

\bibitem[{\textit{Cowley}(1973)}]{cowley73a}
Cowley, S. W.~H. (1973), A quanlitative study of the reconnection between the
  earth's magnetic field and an interplanetary field of arbitrary orientation,
  \textit{Radio Sci.}, \textit{8}, 903--913.

\bibitem[{\textit{Cowley}(1976)}]{cowley76a}
Cowley, S. W.~H. (1976), Comments on the merging of nonantiparallel magnetic
  fields, \textit{J. Geophs. Res}, \textit{81}(19), 3455.

\bibitem[{\textit{Daly et~al.}(1984)\textit{Daly, Saunders, Rijnbeek, Sckopke,
  and Russell}}]{daly84a}
Daly, P.~W., M.~A. Saunders, R.~P. Rijnbeek, N.~Sckopke, and C.~T. Russell
  (1984), The distribution of reconnection geometry in flux transfer events
  using energetic ion, plasma and magnetic data, \textit{J. Geophs. Res},
  \textit{89}, 3843--3854.

\bibitem[{\textit{Daughton et~al.}(2011)\textit{Daughton, Roytershteyn,
  Karimabadi, Yin, Albright, Bergen, and Bowers}}]{daughton11a}
Daughton, W., V.~Roytershteyn, H.~Karimabadi, L.~Yin, B.~J. Albright,
  B.~Bergen, and K.~J. Bowers (2011), Role of electron physics in the
  development of turbulent magnetic reconnection in collisionless plasmas,
  \textit{Nature Physics}, \textit{7}, 539--542, \doi{10.1038/nphys1965}.

\bibitem[{\textit{Denton et~al.}(2012)\textit{Denton, Sonnerup, Swisdak, Birn,
  Drake, and Hesse}}]{denton12a}
Denton, R.~E., B.~U.~{\"O}. Sonnerup, M.~Swisdak, J.~Birn, J.~F. Drake, and
  M.~Hesse (2012), Test of shi et al. method to infer the magnetic reconnection
  geometry from spacecraft data: Mhd simulation with guide field and
  antiparralel kinetic simulation, \textit{J. Geophys. Res.}, \textit{117},
  A09,201.

\bibitem[{\textit{Dorelli et~al.}(2007)\textit{Dorelli, Bhattacharjee, and
  Raeder}}]{dorelli07a}
Dorelli, J.~C., A.~Bhattacharjee, and J.~Raeder (2007), Separator reconnection
  at earth's dayside magnetopause under generic northward interplanetary
  magnetic field conditions, \textit{J. Geophs. Res}, \textit{112}, A02,202.

\bibitem[{\textit{Dungey}(1961)}]{dungey61a}
Dungey, J.~W. (1961), Interplanetary magnetic field and the auroral zone,
  \textit{Phys. Rev. Lett.}, \textit{6}, 47--48.

\bibitem[{\textit{Dunlop et~al.}(2011{\natexlab{a}})}]{dunlop11b}
Dunlop, M.~W., et~al. (2011{\natexlab{a}}), Extented magnetic reconnection
  across the dayside magnetopause, \textit{Phys. Rev. Lett.}, \textit{107},
  025,004.

\bibitem[{\textit{Dunlop et~al.}(2011{\natexlab{b}})}]{dunlop11a}
Dunlop, M.~W., et~al. (2011{\natexlab{b}}), Magnetopause reconnection across
  wide local time, \textit{Ann. Geophys.}, \textit{29}, 1683--1697.

\bibitem[{\textit{Fear et~al.}(2012)\textit{Fear, Palmroth, and
  Milan}}]{fear12a}
Fear, R.~C., M.~Palmroth, and S.~E. Milan (2012), Seasonal and clock angle
  control of the location of flux transfer event signatures at the
  magnetopause, \textit{J. Geophs. Res}, \textit{117}, A04,202.

\bibitem[{\textit{Gonzalez and Mozer}(1974)}]{gonzalez74a}
Gonzalez, W.~D., and F.~S. Mozer (1974), A quantitative model for the potential
  resulting from reconnection with an arbitrary interplanetary magnetic field,
  \textit{J. Geophs. Res}, \textit{79}(28), 4186--4194.

\bibitem[{\textit{Hesse and Birn}(1993)}]{hesse93a}
Hesse, M., and J.~Birn (1993), Parallel electric fields as acceleration
  mechanisms in three-dimensional reconnection, \textit{Adv. Space Res.},
  \textit{13}, 249.

\bibitem[{\textit{Hesse and Schindler}(1988)}]{hesse88a}
Hesse, M., and K.~Schindler (1988), A theoretical foundation of general
  magnetic reconnection, \textit{J. Geophys. Res.}, \textit{93}(A6),
  5559--5567.

\bibitem[{\textit{Hesse et~al.}(2013)\textit{Hesse, Aunai, Zenitani,
  Kuznetsova, and Birn}}]{hesse13a}
Hesse, M., N.~Aunai, S.~Zenitani, M.~Kuznetsova, and J.~Birn (2013), Aspects of
  collisionless magnetic reconnection in asymmetric systems, \textit{Phys.
  Plasmas}, \textit{20}, 061,210.

\bibitem[{\textit{Kawano and Russell}(2005)}]{kawano05a}
Kawano, H., and C.~T. Russell (2005), Dual-satellite observations of the
  motions of flux transfer events: Statistical analysis with isee 1 and isee 2,
  \textit{J. Geophys. Res.}, \textit{110}, A07,217.

\bibitem[{\textit{Komar et~al.}(2013)\textit{Komar, Cassak, Dorelli, Glocer,
  and Kuznetsova}}]{komar13a}
Komar, C.~M., P.~A. Cassak, J.~C. Dorelli, A.~Glocer, and M.~M. Kuznetsova
  (2013), Tracing magnetic separators and their dependence on imf clock angle
  in global magnetospheric simulations, \textit{J. Geophs. Res}, \textit{118},
  4998--5007.

\bibitem[{\textit{Komar et~al.}(2015)\textit{Komar, Fermo, and
  Cassak}}]{komar15a}
Komar, C.~M., R.~L. Fermo, and P.~A. Cassak (2015), The dayside reconnetion x
  line, \textit{J. Geophs. Res}, \textit{120}, 276--294.

\bibitem[{\textit{Moore et~al.}(2002)\textit{Moore, Fok, and
  Chandler}}]{moore02a}
Moore, T.~E., M.~C. Fok, and M.~O. Chandler (2002), The dayside reconnetion x
  line, \textit{J. Geophs. Res}, \textit{107}(A10), 1332.

\bibitem[{\textit{Papadopoulos et~al.}(1999)\textit{Papadopoulos, Goodrich,
  Wiltberger, Lopez, and Lyon}}]{papadopoulos99a}
Papadopoulos, K., C.~Goodrich, M.~Wiltberger, R.~Lopez, and J.~Lyon (1999), The
  physics of substorms as revealed by the istp, \textit{Phys. Chem. Earth Part
  C}, \textit{24}, 189--202.

\bibitem[{\textit{Phan and Paschmann}(1996)}]{phan96a}
Phan, T.~D., and G.~Paschmann (1996), Low-latitude dayside magnetopause and
  boundary layer for high magnetic shear 1. structure and motion, \textit{J.
  Geophys. Res.}, \textit{101}(A4), 7801.

\bibitem[{\textit{Phan et~al.}(2006)\textit{Phan, Hasegawa, Fujimoto, Oieroset,
  Mukai, Lin, and Paterson}}]{phan06b}
Phan, T.~D., H.~Hasegawa, M.~Fujimoto, M.~Oieroset, T.~Mukai, R.~P. Lin, and
  W.~Paterson (2006), Simultaneous geotail and wind observations of
  reconnection at the subsolar and tail flank magnetopause, \textit{Geophys.
  Res. Lett.}, \textit{33}, L09,104.

\bibitem[{\textit{Phan et~al.}(2010)\textit{Phan, Gosling, Paschmann, Pasma,
  Drake, {\O}ieroset, Larson, Lin, and Davis}}]{phan10a}
Phan, T.~D., J.~T. Gosling, G.~Paschmann, C.~Pasma, J.~F. Drake,
  M.~{\O}ieroset, D.~Larson, R.~P. Lin, and M.~S. Davis (2010), The dependence
  of magnetic reconnection on plasma $\beta$ and magnetic shear: Evidence from
  solar wind observations, \textit{Astrophys. J.}, \textit{719}, L199--L203,
  \doi{10.1088/2041-8205/719/2/L199}.

\bibitem[{\textit{Pritchett}(2008)}]{pritchett08a}
Pritchett, P.~L. (2008), Collisionless magnetic reconnection in an asymmetric
  current sheet, \textit{J. Geophys. Res.}, \textit{113}, A06,210.

\bibitem[{\textit{Pu et~al.}(2007)}]{pu07a}
Pu, Z.~Y., et~al. (2007), Global view of dayside magnetic reconnection with the
  dusk-dawn imf orientation: A statistical study for double star and cluster
  data, \textit{Geophys. Res. Lett.}, \textit{34}, L20,101.

\bibitem[{\textit{Schindler et~al.}(1988)\textit{Schindler, Hesse, and
  Birn}}]{schindler88a}
Schindler, K., M.~Hesse, and J.~Birn (1988), General magnetic reconnection,
  parallel electric fields, and helicity, \textit{J. Geophys. Res.},
  \textit{93}(A6), 5547--5557.

\bibitem[{\textit{Schreier et~al.}(2010)\textit{Schreier, Swisdak, Drake, and
  Cassak}}]{schreier10a}
Schreier, R., M.~Swisdak, J.~F. Drake, and P.~A. Cassak (2010),
  Three-dimensional simulations of the orientation and structure of
  reconnection x-lines, \textit{Phys. Plasmas}, \textit{17}, 110,704.

\bibitem[{\textit{Scurry et~al.}(1994)\textit{Scurry, Russell, and
  Gosling}}]{scurry94a}
Scurry, L., C.~T. Russell, and J.~T. Gosling (1994), A statistical study of
  accelerated flow events at the dayside magnetopause, \textit{J. Geophs. Res},
  \textit{99}(A8), 14,815--14,829.

\bibitem[{\textit{Shay et~al.}(1999)\textit{Shay, Drake, Rogers, and
  Denton}}]{shay99a}
Shay, M.~A., J.~F. Drake, B.~N. Rogers, and R.~E. Denton (1999), The scaling of
  collisionless, magnetic reconnection for large systems, \textit{Geophys. Res.
  Lett.}, \textit{26}, 2163--2166.

\bibitem[{\textit{Shay et~al.}(2003)\textit{Shay, Drake, Swisdak, Dorland, and
  Rogers}}]{shay03a}
Shay, M.~A., J.~F. Drake, M.~Swisdak, W.~Dorland, and B.~N. Rogers (2003),
  Inherently three dimensional magnetic reconnection: A mechanism for bursty
  bulk flows?, \textit{Geophys. Res. Lett.}, \textit{30}(6), 1345.

\bibitem[{\textit{Shi et~al.}(2005)\textit{Shi, Shen, Pu, Dunlop, Zhang, Zhang,
  Xiao, Liu, and Balogh}}]{shi05a}
Shi, Q.~Q., C.~Shen, Z.~Y. Pu, M.~W. Dunlop, Q.~G. Zhang, H.~Zhang, C.~J. Xiao,
  Z.~X. Liu, and A.~Balogh (2005), Dimensional analysis of observed structures
  using multipoint magnetic field measurements: Application to cluster,
  \textit{Geophys. Res. Lett.}, \textit{32}, L12,105.

\bibitem[{\textit{Sibeck}(2009)}]{sibeck09a}
Sibeck, D.~G. (2009), Concerning the occurence pattern of flux transfer events
  on the dayside magnetopause, \textit{Ann. Geophys.}, \textit{27}, 895--903.

\bibitem[{\textit{Siscoe et~al.}(2001)\textit{Siscoe, Erickson, Sonnerup,
  Maynard, Siebert, Weimer, and White}}]{siscoe01a}
Siscoe, G.~L., G.~M. Erickson, B.~U.~{\"O}. Sonnerup, N.~C. Maynard, K.~D.
  Siebert, D.~R. Weimer, and W.~W. White (2001), Global role of $e_\|$ in
  magnetopause reconnecgtion: An explicit demonstration, \textit{J. Geophs.
  Res}, \textit{106}, 13,015--13,022.

\bibitem[{\textit{Sonnerup}(1974)}]{sonnerup74a}
Sonnerup, B. U.~{\"O}. (1974), Magnetopause reconnection rate, \textit{J.
  Geophys. Res.}, \textit{79}(10), 1546--1549.

\bibitem[{\textit{Stanier et~al.}(2015)\textit{Stanier, Simakov, Chaco\'n, and
  Daughton}}]{stanier15a}
Stanier, A., A.~N. Simakov, L.~Chaco\'n, and W.~Daughton (2015), Fluid vs.
  kinetic magnetic reconnection with strong guide fields, \textit{submitted to
  Phys. Plasmas}.

\bibitem[{\textit{Swisdak and Drake}(2007)}]{swisdak07a}
Swisdak, M., and J.~F. Drake (2007), Orientaion of the reconnection x-line,
  \textit{Geophys. Res. Lett.}, \textit{34}, L11,106.

\bibitem[{\textit{Swisdak et~al.}(2010)\textit{Swisdak, Opher, Drake, and
  Bibi}}]{swisdak10a}
Swisdak, M., M.~Opher, J.~F. Drake, and F.~A. Bibi (2010), The vector direction
  of the interstellar magnetic field outside the heliosphere,
  \textit{Astrophys. J.}, \textit{710}, 1769--1775.

\bibitem[{\textit{Teh and Sonnerup}(2008)}]{teh08a}
Teh, W.~L., and B.~U.~{\"O}. Sonnerup (2008), First results from ideal 2-d mhd
  reconstruction: Magnetopause reconnection event seen by cluster, \textit{Ann.
  Geophys.}, \textit{26}(9), 2673--2684.

\bibitem[{\textit{Trattner et~al.}(2007)\textit{Trattner, Mulcock, Petrinec,
  and Fuselier}}]{trattner07a}
Trattner, K.~J., J.~S. Mulcock, S.~M. Petrinec, and S.~A. Fuselier (2007),
  Probing the boundary between antiparallel and component reconnection during
  souwthward interplanetary magneitc field conditions, \textit{J. Geophs. Res},
  \textit{112}, A01,201.

\bibitem[{\textit{Wild et~al.}(2007)}]{wild07a}
Wild, J.~A., et~al. (2007), On the location of dayside magnetic reconnection
  during an interval of duskward oriented imf, \textit{Ann. Geophys.},
  \textit{25}, 219--238.

\bibitem[{\textit{{Yi-Hsin Liu} et~al.}(2013)\textit{{Yi-Hsin Liu}, Daughton,
  Karimabadi, Li, and Roytershteyn}}]{yhliu13a}
{Yi-Hsin Liu}, W.~Daughton, H.~Karimabadi, H.~Li, and V.~Roytershteyn (2013),
  Bifurcated structure of the electron diffusion region in three-dimensional
  magnetic reconnection, \textit{Phys. Rev. Lett.}, \textit{110}, 265,004.

\bibitem[{\textit{{Yi-Hsin Liu} et~al.}(2014)\textit{{Yi-Hsin Liu}, Daughton,
  Karimabadi, Li, and Gary}}]{yhliu14a}
{Yi-Hsin Liu}, W.~Daughton, H.~Karimabadi, H.~Li, and S.~P. Gary (2014), Do
  dispersive waves play a role in collisionless magnetic reconnection?,
  \textit{Phys. Plasmas}, \textit{21}, 022,113.

\end{thebibliography}

\end{article}

\clearpage


\begin{table}[ht]
\caption{Parameters of Runs } 
\centering 
\begin{tabular}{c c c c c c c c c c} 
\hline\hline 
Cases & Type &$m_i/m_e$ & $\lambda/d_i$ & $L_x/d_i$ & $L_y/d_i$ & $L_z/d_i$ & $\theta_{box}$ & $\theta_{xline}$ \\ [0.5ex] 
\hline 
a & 3D & 1 & 2.5 & 256 & 256 & 32  & $10^\circ$ & $\approx -28^\circ$ \\
b & 3D & 1 & 1.5 & 128 & 128 & 25.6  & $0^\circ$ & $\approx -28^\circ$ \\
c & 2D & 1 & 2.5 & 256 & NA & 32 & $[10^\circ, -37^\circ]$ & $\theta_{box}$ \\
d & tearing & 1 & 1.36 & 256 & 256 & 32  & $10^\circ$ & NA \\
e & 3D & 4 & 1.5 & 128 & 128 & 25.6 &  $0^\circ$& $\approx -21^\circ$ \\
f & 2D & 4 & 1.5 & 128 & NA & 25.6 &  $[10^\circ, -34^\circ]$ & $\theta_{box}$\\
g & 3D & 4 & 1.5 & 83.2 & 83.2 & 25.6 &  $0^\circ$& $\approx -21^\circ$ \\
h & 3D & 4 & 1.5 & 64 & 64 & 25.6 &  $0^\circ$& $\approx -21^\circ$ \\
i & 3D & 4 & 1.5 & 64 & 83.2 & 25.6 &  $0^\circ$& $\approx -21^\circ$ \\
j & 3D &4 & 1.5 & 64 & 83.2 & 25.6 &  $-15^\circ$& $\approx -21^\circ$ \\
k & 3D & 25 & 0.8 & 64 & 64 & 16 & $10^\circ$ & $\approx -13^\circ$\\
l & 3D & 25 & 0.8 & 64 & 64 & 16 & $0^\circ$ & $\approx -13^\circ$\\
m & 2D & 25 & 0.8 & 64 & NA & 16 & $[10^\circ, -33^\circ]$ & $\theta_{box}$\\
n & tearing & 25 & 0.272 & 64 & 64 & 16 & $10^\circ$ & NA \\
o & 2D & 256 & 0.8 & 64 & 64 & 16 & $[10^\circ,-33^\circ]$  & $\theta_{box}$\\
\hline
\end{tabular}
\label{table:nonlin} 
\end{table}

\newpage


\end{document}